\documentclass[preprint,aps,prc]{revtex4}
\usepackage{amsmath}  
\usepackage{epsfig} 
\usepackage{color} 
\usepackage{txfonts}  
 
\begin{document}

\parindent=12pt

\parindent=12pt

\title{A derivation of   transition density from the observed 
\boldmath{ $^4{\rm He}(e,e') ^4{\rm He}(0_2^+)$} \\ 
form factor  raising the $\alpha$-particle monopole puzzle}

\author{M.\ Kamimura}
\email{mkamimura@a.riken.jp}
\affiliation{Meson Science Laboratory, RIKEN Nishina Center, RIKEN, Wako 351-0198, Japan}

\date{\today}

\begin{abstract}%
Recently, the monopole transition form factor of the electron-scattering 
excitation of the $0^+_2$ state ($E_x=20.21$ MeV) of the $^4{\rm He}$ nucleus was
observed over a broad momentum transfer range 
($ 0.5 \leq q^2 \leq 5.0 \,{\rm  fm}^{-2}$)
with dramatically improved preision compared with older sets of data;
modern nuclear forces, including those derived 
from the chiral effective field theory,
failed to reproduce the form factor, which is called \mbox{$\alpha$-particle} 
monopole puzzle.  
To resolve this puzzle by improving the study of spatial 
structure of the $0^+_2$ state, we derive in this letter 
a possible $0^+_1\!\to\! 0^+_2$ transition density $\rho_{\rm  tr}(r)$
for \mbox{$r \gtrsim 1$ fm} from the observed form factor.
The shape of the transition density is 
significantly different from that obtained theoretically in the literature.
\end{abstract}

\maketitle

\noindent
{\bf 1. Introduction}
 
The $^4$He nucleus is the lightest nucleus that exhibits excited
states (resonances).
One of the important tasks in few-body nuclear physics is to
solve the four-nucleon states as a stringent test for
{\it ab initio} few-body methods and nuclear Hamiltonian.
A benchmark test calculation for this purpose was reported in
Ref.~\cite{Kama01} (2001) by  seven different few-body research
groups. The $^4$He ground state was solved  using a realistic
interaction, the Argonne V8$^\prime$ potential 
(AV8$'$)~\cite{AV8P97}.   
 Agreement between 
the results of the significantly different calculational
schemes was essentially perfect in terms of the binding energy, the
root mean square radius, and the two-body correlation function.
The present author participated in the benchmark test 
together with Hiyama, using the Gaussian expansion method (GEM) for few-body 
systems~\cite{Kamimura88,Kameyama89,Hiyama03}.

One of the next challenging projects was to explain the properties of the second
$0^+$ state \mbox{(a resonance)} of $^4$He using realistic interactions, 
simultaneously reproducing the $0^+_1$  and $0^+_2$  states
with significantly different spatial structures.
Hiyama, Gibson, and the present author~\cite{second2004} (2004),
employed the GEM and the AV8$'$  + phenomenological  central 3N potential
to reproduce the binding energies 
of $^3$H, $^3$He and $^4$He($0^+_1)$, and predicted the energy of
$^4$He($0^+_2)$ as $-8.19$ MeV measured from the four-body breakup threshold
without any additional adjustable parameters
on the basis of the bound-state
approximation with approximately 100 four-body angular momentum channels
in the isospin formalism.
The calculated energy of the $0^+_2$ state with isospin $T\!=\!0$ explained the 
observed value  $-8.09$ MeV with a 100 keV error.
The calculated transition form factor of $^4{\rm He}(e,e') ^4{\rm He}(0_2^+)$
agreed with  available data~\cite{Fro68,Wal70,Koe83}
within  large observed errors (Fig.~3~\cite{second2004}). 

It was further found that (i) the percentage probabilities of the 
$S$-, $P$- and $D$-components in
the $0^+_2$ state are almost the same as those of $^3$H and $^3$He,
and (ii) the overlap amplitude  between the 
$^4{\rm He}(0^+_n)$ wave function
and the $^3{\rm H}$ wave function (see Fig.~4 of Ref.~\cite{second2004})
represents that, in the ground state, the fourth nucleon is located close
to the other three nucleons, but it is far away from them in
the second 0+ state.
These analyses 
indicate that the second $0^+$ state has  well-developed $3N+N$ 
cluster structure with relative $S$-wave motion, not having a monopole breathing mode. 
This state property  was soon confirmed by Horiuchi and Suzuki~\cite{Horiuchi}
adopting the stochastic variational method with 
the correlated Gaussian basis functions~\cite{stochas-1,stochas-2} which was
one of the numerical methods 
used for \mbox{the benchmark test~\cite{Kama01}}.

Bacca {\it et al.}~\cite{Bacca2013,Bacca2015} investigated 
the second $0^+$ state
using the effective-interaction hyperspherical
harmonic (EIHH) 
method~\cite{EIHH2001,EIHH1997}, which is  also one of the methods 
for the benchmark calculation~\cite{Kama01}.
Further, they  employed the Lorentz integral transformation
(LIT) approach~\cite{LIT1994,LIT2007} for the calculation of resonant states.
For Hamiltonians, they used 
(i) Argonne $V_{18}$ (AV18)~\cite{AV18} potential plus
Urbana IX (UIX)~\cite{UIX} 3NF and 
(ii) a chiral effective field theory ($\chi$EFT) based potential 
(N$^3$LO NN~\cite{EFT-NN} +  N$^2$LO 3NF~\cite{EFT-3FN-1,EFT-3FN-2}). 
They found that the calculated results of the transition form factor
are strongly dependent on the Hamiltonian and do not agree with the
experimental data~\cite{Fro68,Wal70,Koe83}, especially in the case of   
the $\chi$EFT potential.
The authors claimed that it was highly desirable 
to have a further experimental 
confirmation of the existing data and, in particular, with increased precision.

In response, a new experiment on the transition form factor 
with significantly improved precision
was performed at the Maintz Microtoron by Kegel {\it et al.}~\cite{Kegel}.
All the four authors of Refs.~\cite{Bacca2013,Bacca2015}
participated in this work~\cite{Kegel} as co-authors.
The precise results for the form factor are shown in
Figs.~3 and 4~\cite{Kegel}, and
confirm  previous data~\cite{Fro68,Wal70,Koe83} with
much higher precision. 
They found that the {\it ab initio} calculations~\cite{second2004,Bacca2013}
disagree with the observed form factors; 
for example, the $\chi$EFT result~\cite{Bacca2013} is 100\% too high
with respect to the new data  at the peak position. 

The authors of Ref.~\cite{Kegel} noticed that, in the momentum transfer range
$ 0.2 \leq q^2 \leq 1 \, {\rm fm}^{-2}$, the simplified potential 
in Ref.~\cite{second2004} leads to agreement with the experimental data,   
whereas the realistic calculations~\cite{Bacca2013} do not 
(Fig.~4~\cite{Kegel}).
They showed that the difference did not stem from the numerical
methods but from the Hamiltonian; to examine it, they employed the same potential 
(AV8$'$ + central 3N) of Ref.~\cite{second2004} using their calculation 
method (EIHH) and reproduced the result of Ref.~\cite{second2004} 
as compared in Fig.~4 of Ref.~\cite{Kegel}.

Regarding the explicit information on the spatial structure of the $0^+_2$ state,
the two gross features of the transition density, 
monopole matrix element $\langle r^2 \rangle_{\rm tr}$ and 
transition radius $\mathcal{R}_{\rm tr}$,  were extracted 
based on the behavior of the form factor at $q \sim 0$. 
It was noticed~\cite{Kegel} that the AV8$'$ + central 3N potential 
is not compatible with the experimental value of 
$\langle r^2 \rangle_{\rm tr}$, while the realistic AV18 + UIX 
fits the value, and the $\chi$EFT potential prediction deviates the most 
from the experiments even at low momentum values. 
Further discussion on  $\langle r^2 \rangle_{\rm tr}$ and 
$\mathcal{R}_{\rm tr}$ is made in Sec.~4 in this letter.

In  Ref.~\cite{Kegel}, it was concluded  that
there is a puzzle that is not caused by the applied few-body methods,
but rather by the modeling of the nuclear Hamiltonian, and therfore
further theoretical work is needed to resolve the
$\alpha$-particle monopole puzzle.  
On the day of publication of 
Ref.~\cite{Kegel}, this puzzle was introduced 
and discussed in a review article by Epelbaum~\cite{Epelbaum};
Fig.~2 in Ref.~\cite{Epelbaum} summarizes the  transition form factors
obsereved newly by Ref.~\cite{Kegel} and previously by 
\mbox{Refs.~\cite{Fro68,Wal70,Koe83}},
and values calculated in Ref.~\cite{second2004} (yellow line)
and Ref.~\cite{Bacca2013} (blue and red lines).

More information regarding the spatial structure of the second $0^+$ state 
is required to solve this puzzle.
Thus, the purpose of this letter is to 
extract a possible mass transition density from the newly
observed $^4{\rm He}(e,e') ^4{\rm He}(0_2^+)$ form factor.

\vskip 0.3cm
\noindent
{\bf 2. Transition form factor and transition density}

To discuss the mass transition density, we introduce the  
mass form factor ${\widetilde {\cal F}}_{{\rm M}0^+}(q^2)$:
\begin{eqnarray}
      |{\widetilde {\cal F}}_{{\rm M}0^+}(q^2)|^2 = 
         |\mathcal{F}_{{\rm M}0^+}(q^2)|^2/f_p(q^2) ,
\label{eq:mass-FF}
\end{eqnarray}
where $f_p(q^2)=(1/(1+0.0548 q^2))^4$ denotes the proton finite-size
factor ($q$ in fm$^{-1})$.
In Fig.~1, \mbox{we illustrate} the observed data~\cite{Kegel} of 
$|{\cal F}_{{\rm M}0^+}(q^2)|^2$
in the form of  $|{\widetilde {\cal F}}_{{\rm M}0^+}(q^2)|/q^2$
by the open circles; 
\mbox{the relative error} in the observed  $|{\cal F}_{{\rm M}0^+}(q^2)|^2$  is 
$4\%\!-\!6\%$ except for 11\% at the \mbox{lowest $q^2$}. 
We take an approximation of $Q^2=q^2$ for the four-momentum transfer $Q^2$
when using Fig.3 of Ref.~\cite{Kegel}.
\footnote{   
In Ref.~\cite{Kegel}, 
the data presented in Fig.~3 are shown as  functions of 
the four-momentum transfer $Q^2$
given by $Q^2=q^2 - (\hbar \omega/\hbar c)^2$
with $q=|{\bf q}|$ as the three-momentum transfer and
$\hbar \omega \,(0^+_2) =20.2 $MeV \mbox{(cf. page 2} of Ref.~\cite{Kegel}), 
whereas low-momentum data shown in Fig.~4 are provided with respect to $q^2$.
We approximate $Q^2=q^2$ considering $Q^2=q^2 - 0.01\,{\rm fm}^{-2}$
and  heavy use of Fourier transformation with  $ e^{i{\bf q}\cdot{\bf r} }$
such as in Eqs.~(5) and (6). 
}  

\begin{figure} [h!]
\begin{center} 
\epsfig{file=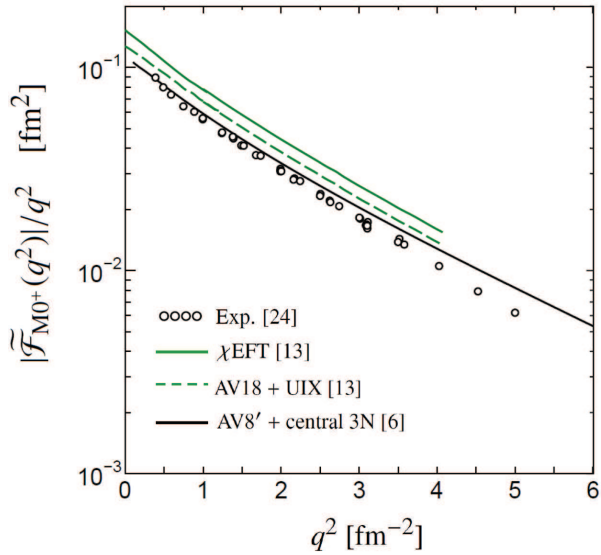,width=8.0cm}
\caption{ Transition mass form factor $|{\widetilde {\cal F}}_{{\rm M}0^+}(q^2)|$
divided by $q^2$. The observed data 
are shown by the open circles which are 
transformed from the open circles for $|{\cal F}_{{\rm M}0^+}(q^2)|^2$ given  
in Figs.~3 and 4 of Ref.~\cite{Kegel};
the relative error in $|{\cal F}_{{\rm M}0^+}(q^2)|^2$ is 
$4\%\!-\!6\%$ except for 11\% at the lowest $q^2$. 
The black line represents the calculated form factor provided 
by Hiyama {\it et al.}~\cite{second2004}
using the AV8$'$+central 3N potential. 
The green line (green dashed line) was derived by  
Bacca {\it et al.}~\cite{Bacca2013}  using the $\chi$EFT (AV18+UIX) 
potential; the lines are transformed from the corresponding lines 
in Figs.~3 and 4 of Ref.~\cite{Kegel}.
} 
\label{fig:FF-fig1}
\end{center}	
\end{figure}

The black line represents the form factor provided 
by Hiyama {\it et al.}~\cite{second2004}
using the AV8$'$+central 3N potential.  Bacca {\it et al.}~\cite{Bacca2013}
derived the green line (green dashed line)  
using the $\chi$EFT (AV18+UIX) potential;
the lines are transformed from the corresponding lines 
in Figs.~3 and 4 of Ref.~\cite{Kegel}.

In Fig.~2, the observed data are simulated by the blue line which is 
transformed from the blue line in Figs.~3 and 4 of Ref.~\cite{Kegel}.
By virtue of the semi-log plot of Fig.~2, we find that 
the blue line $(0 \leq q^2 \leq 5.5\,{\rm fm}^{-2})$ is well simulated 
by a sum of two ``straight lines'', namely,
%
\begin{eqnarray}
{\widetilde {\cal F}}_{{\rm M}0^+}(q^2)/q^2 = a_1 e^{-b_1 q^2}
                   +a_2 \,e^{-b_2 q^2} 
\label{eq:approx-FF}
\end{eqnarray}
with $(a_1,\, a_2)=(0.092, \,0.035)$ fm$^{2}$
and $(b_1,\, b_2)=(0.54,\, 2.5)$ fm$^{2}$, which  is illustrated
in Fig.~2  by the red line; 
here we fitted the parameters to two significant figures
taking the error of the observed data into  consideration.
This makes it possible to derive a transition mass density 
from the observed transition form factor as is discussed below. 

\begin{figure} [b!]
\begin{center} 
\epsfig{file=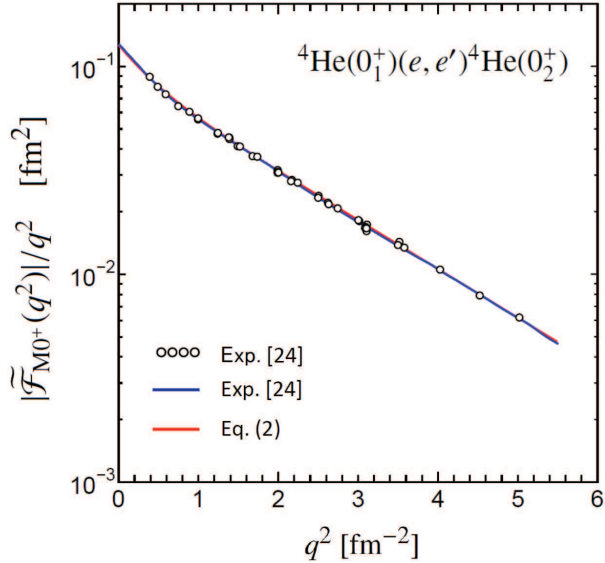,width=8.0cm}
\caption{ Transition mass form factor $|{\widetilde {\cal F}}_{{\rm M}0^+}(q^2)|$
divided by $q^2$. The observed data~\cite{Kegel} 
are shown by the open circles that are the same as those in Fig.~1.
The data are  fitted by the blue line, which is transformed from
the blue line (the spline polynomial fit function) in Figs.~3 and 4 
of Ref.~\cite{Kegel}, whereas the red line shows 
$|{\widetilde {\cal F}}_{{\rm M}0^+}(q^2)|/q^2$ approximated by Eq.~(2).
} 
\label{fig:FF-fig2}
\end{center}	
\end{figure}

Monopole mass transition density $\rho_{\rm tr}(r)$ is defined as
\begin{eqnarray}
\rho_{\rm tr}(r)= \langle 0^+_2\,| \frac{1}{4} \sum_{i=1}^{4}
\delta({\bf r}-{\bf r}_i)|\,0^+_1 \rangle,
\end{eqnarray}
where the $|\,0^+_1 \rangle$ and $|\,0^+_2 \rangle$ are the
four-nucleon wave functions (cf.  Eq.~(2.1) in Ref.~\cite{second2004}),
and ${\bf r}_i$ is the position vector of $i$th nucleon with respect to
the center-of-mass of the four nucleons.
This definition is the same as that in Bacca {\it et al.}~\cite{Bacca2015}
in their Eq.~(9) and Fig.~1.
We have 
\begin{eqnarray}
\int \, \rho_{\rm tr}(r)\, {\rm d}{\bf r}=0 
\end{eqnarray}
because the left hand side is the overlap 
$\langle 0^+_2\,|\,0^+_1 \rangle$ between the $0^+_1$ and $0^+_2$ states.

The form factor ${\widetilde {\cal F}}_{{\rm M}0^+}(q^2)$ is related to the
monopole transition density $\rho_{\rm tr}(r)$ as follows:
\begin{eqnarray}
{\widetilde {\cal F}}_{{\rm M}0^+}(q^2) =
\int e^{i{\bf q}\cdot{\bf r}} \, \rho_{\rm tr}(r)\, {\rm d}{\bf r}.
\label{eq:dens-to-FF}
\end{eqnarray}
The density $\rho_{\rm tr}(r)$ can be derived as 
\begin{eqnarray}
\rho_{\rm tr}(r) =
\frac{1}{(2\pi)^3}
\int e^{-i{\bf q}\cdot{\bf r}} \,{\widetilde {\cal F}}_{{\rm M}0^+}(q^2)
  \,{\rm d}{\bf q} 
\label{eq:FF-to-dens}
\end{eqnarray}
if ${\widetilde {\cal F}}_{{\rm M}0^+}(q^2)$ is provided  
for all $q^2$-range.

\begin{figure} [b!]
\begin{center}
\epsfig{file=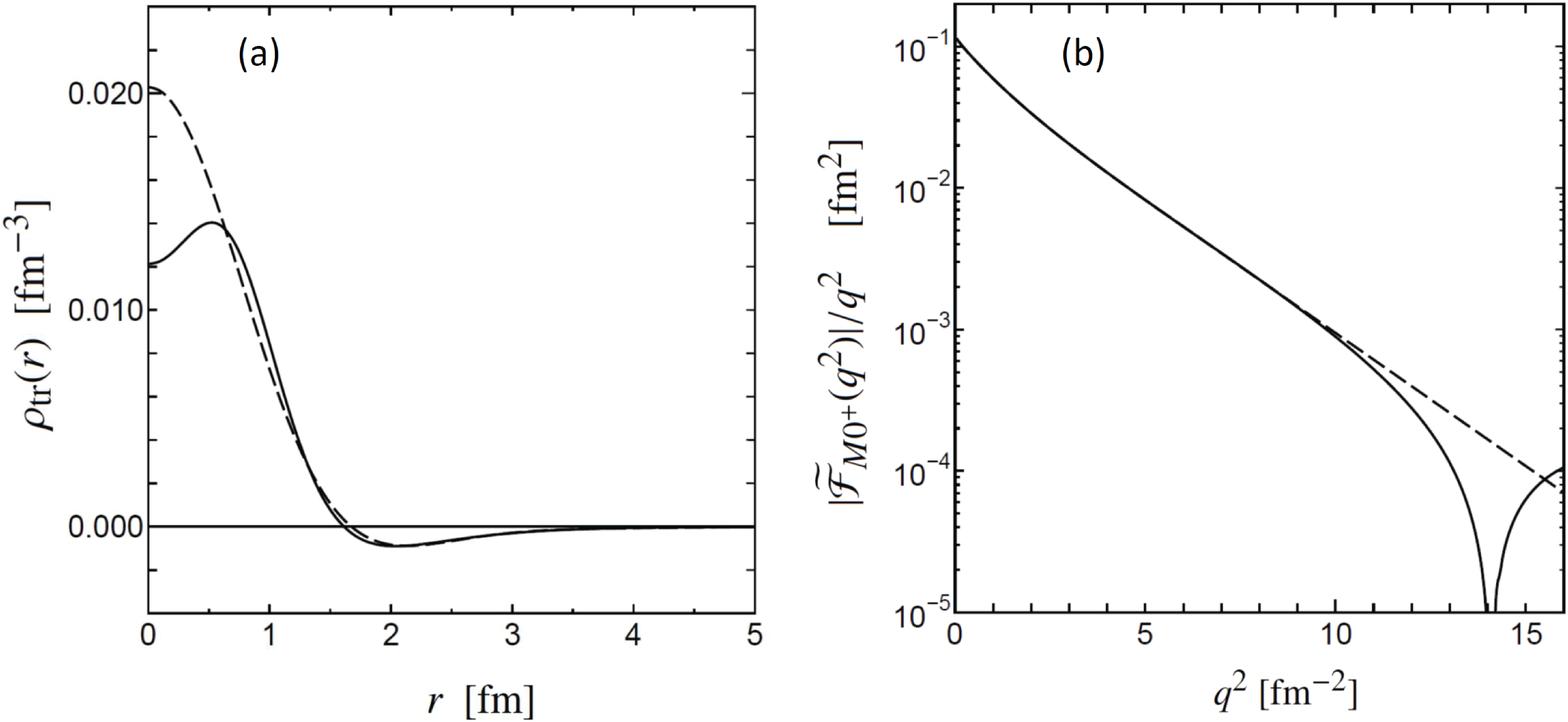,width=15.0cm}
\caption{
The solid line in (a) illustrates 
the calculated mass transition mass density $\rho_{\rm tr}(r)$
in Ref.~\cite{second2004}, 
which yields the solid-line form factor in (b).
The dashed-line form factor in (b) generates
the dashed-line density with no central depression in (a).
} 
\label{fig:dens-fig3ab}
\end{center}	
\end{figure}

\begin{figure} [b!]
\begin{center}
\epsfig{file=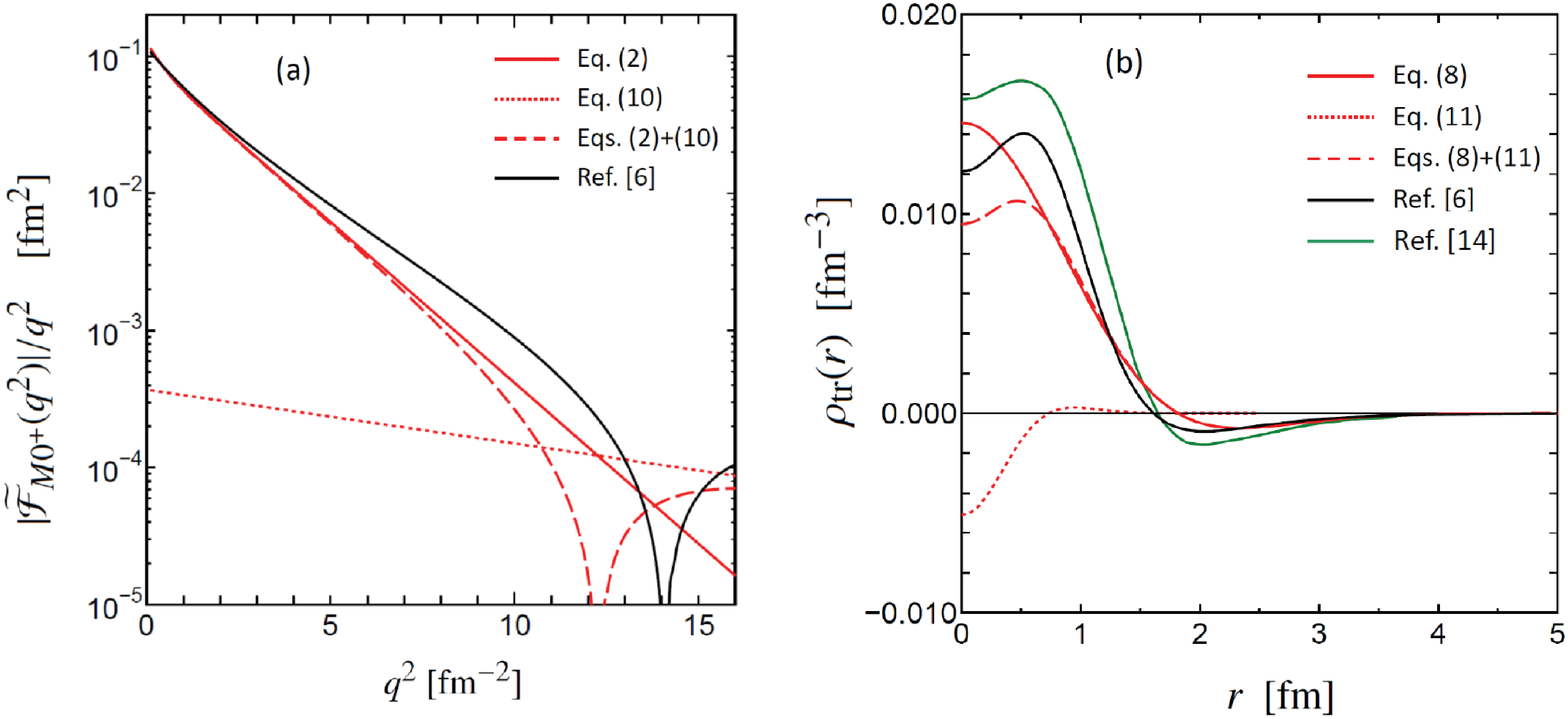,width=15.0cm}
\caption{ 
(a) The red solid line is the form factor given by Eq.~(2). 
The red dotted line is the absolute value of example additional form factor Eq.~(10)
for the central depression of the transition density. 
The red dashed line denotes the example form factor, Eq.~(2) plus Eq.~(10),
generating  the central depression.
The black line is the form factor
given by Ref.~\cite{second2004}. 
(b) The red solid line denotes the transition density Eq.~(8).
The red dotted line represents the example additional density Eq.~(11) 
for simulating  central depression.
The red dashed line denotes  the example density, Eq.~(8) plus Eq.~(11), 
having central depression.  The black  line is the density
given in Fig.~2b of Ref.~\cite{second2004}.
The green line denotes the density calculated using the $\chi$EFT potential
given in Fig.~1 of Ref.~\cite{Bacca2015}.
} 
\label{fig:FF-fig4ab}
\end{center}	
\end{figure}

We examined whether extending Eq.~(\ref{eq:approx-FF}) 
to the range of $q^2 > 5.5$ fm$^{-2}$ is meaningful.
As a test example, we employed  $\rho_{\rm tr}(r)$ 
and  ${\widetilde {\cal F}}_{{\rm M}0^+}(q^2)$ 
calculated in Ref.~\cite{second2004}. 
The density $\rho_{\rm tr}(r)$ 
$(=\rho_{00}(r)/\sqrt{4\pi}$~\cite{second2004})
is illustrated in Fig.~3a by the solid line with a node at $r=1.62$ fm, whereas 
$|{\widetilde {\cal F}}_{{\rm M}0^+}(q^2)|/q^2$ 
is shown in Fig.~3b as a solid line with a node at $q^2=14.1\, {\rm fm}^{-2}$
$({\widetilde {\cal F}}_{{\rm M}0^+}(q^2)= \sqrt{4\pi} F_{\rm inel}(q^2) $~\cite{second2004}).
The solid-line form factor in  the range of $3 < q^2 <5.5 \,{\rm fm}^2$ 
extends  to \mbox{$5.5 < q^2 < 9 \,{\rm fm}^2$} to maintain similar decay rates.  
We approximated the solid-line form factor, as shown in Fig.~3b in the range of 
$3 < q^2 < 9 \,{\rm fm}^{-2}$ using the function
\begin{eqnarray}
{\widetilde {\cal F}}_{{\rm M}0^+}(q^2)/q^2 = a_0\, e^{-b_0 q^2}
\end{eqnarray}
with $a_0=0.078 \, {\rm fm}^{2}$ and $b_0=0.45 \, {\rm fm}^2$,
and extended it to the range $q^2 > 9 \,{\rm fm}^{-2}$, as indicated
by the dashed line in Fig.~3b;  in the range of 
$q^2 < 9 \,{\rm fm}^{-2}$,  the dashed line
was considered the same as the solid line. 
We substitute the dashed-line form factor  
into Eq.~(\ref{eq:FF-to-dens}) and obtain the dashed-line
density in Fig.~3a with a node at $r=1.66$ fm. 
The difference between the solid- and
dashed-line form factors in the range of $q^2 > 9 \,{\rm fm}^{-2}$ in Fig.~3b
appears as a {\it central depression} at $r \lesssim 1$ fm  on the solid-line
density in Fig.~3a, and generates only a small difference in the density
for \mbox{$r \gtrsim 1$ fm.} 
Subsequently, we considered that an extension of Eq.~(\ref{eq:approx-FF})
to the range of $q^2 > 5.5$ fm$^{-2}$ 
will be useful for studying the observed form factor.  

We extend the form factor Eq.~(\ref{eq:approx-FF})
in the range of $q^2 > 5.5 \,{\rm fm}^{-2}$, as  illustrated in Fig.~4a 
by the red solid line.
Using the form factor in Eq.~(\ref{eq:approx-FF}) over the entire $q^2$ range, 
we obtain
\begin{eqnarray} 
\rho_{\rm tr}(r) = 
   A_1 \left(3- \tfrac{1}{2b_1}r^2 \right)\, e^{-\frac{1}{4b_1}r^2} 
 + A_2 \left(3- \tfrac{1}{2b_2}r^2\right)\,  e^{-\frac{1}{4b_2}r^2}  
\end{eqnarray}
where
\begin{eqnarray}
A_1=\frac{a_1}{(2\pi)^3} \frac{1}{2b_1} \left(\frac{\pi}{b_1}\right)^{3/2},
 \qquad
A_2=\frac{a_2}{(2\pi)^3} \frac{1}{2b_2} \left(\frac{\pi}{b_2}\right)^{3/2} 
\end{eqnarray}
with $A_1=4.8\times 10^{-2} \;{\rm fm}^{-3}$ and $A_2=4.0 \times 10^{-5}\;{\rm fm}^{-3}$.
The density $\rho_{\rm tr}(r)$ of Eq.~(8) is shown in Fig.~4b
using the solid red line with a node at \mbox{$r=1.82$ fm}, which is
almost the same node position as that of the dominant first term in Eq.~(8)
at $r=\!\sqrt{6 b_1}=1.80$ fm.
For comparison, we show in Fig.~4b the
transition density given by Hiyama {\it et al.}~\cite{second2004} 
by the black line calculated using the AV8$'$ + central 3N potential
and that given by Bacca {\it et al.}~\cite{Bacca2015}  
by the green line calculated using the $\chi$EFT interaction
(extracted from Fig.~1~\cite{Bacca2015}). 

\vskip 0.3cm
\noindent
{\bf 3. Central depression of transition density}

We focus on the central depression of the transition density 
$\rho_{\rm tr}(r)$  in Fig.~4b 
in the range of $r \lesssim 1$ fm
as indicated by the black line~\cite{second2004} 
and the green line~\cite{Bacca2015}.
As shown above, the red-solid-line density with no central depression
in Fig.~4b is generated by  the red-solid-line form factor in Fig.~4a.
We then  attempt to provide an artificial example of a central depression to the
red-solid-line density. 
We considered a small additional transition form factor
$\Delta {\widetilde {\cal F}}_{{\rm M}0^+}(q^2)$,
\begin{eqnarray}
\Delta {\widetilde {\cal F}}_{{\rm M}0^+}(q^2)/q^2
= a_3  \,e^{-b_3 q^2} ,
\end{eqnarray}
which is related to an additional transition density 
$\Delta \rho_{\rm tr}(r)$
(note $\int\! \Delta \rho_{\rm tr}(r)\, {\rm d} {\bf r}=0$)
\begin{eqnarray}
\Delta \rho_{\rm tr}(r)= 
A_3 \left(3- \tfrac{1}{2b_3}r^2\right)\,e^{-\tfrac{1}{4b_3}r^2}
\end{eqnarray}
with 
\begin{eqnarray}
A_3=\frac{a_3}{(2\pi)^3} \frac{1}{2b_3} \left(\frac{\pi}{b_3}\right)^{3/2}.
\end{eqnarray}
We consider, as an example, $a_3=-0.00037$ \,fm$^{2}$ and $b_4=0.090\, {\rm fm}^2$, 
yielding $A_4=-0.0017\, {\rm fm}^{-3}$.
$\Delta {\widetilde {\cal F}}_{{\rm M}0^+}(q^2)/q^2$ and 
$\Delta \rho_{\rm tr}(r)$ are presented  in Fig.~4a and 4b, respectively,
indicated by the red dotted line.
The summed form factor
$({\widetilde {\cal F}}_{{\rm M}0^+}(q^2)+
\Delta {\widetilde {\cal F}}_{{\rm M}0^+}(q^2))/q^2$ is 
shown in Fig.~4a by a red dashed line with a node at $q^2=12.2 \,{\rm fm}^{-2}$.
The transition density $\rho_{\rm tr}(r)+ \Delta \rho_{\rm tr}(r)$ is 
illustrated  in Fig.~3b by the red dashed line, which is close to 
the red-solid line $\rho_{\rm tr}(r)$ within the range of
$r \gtrsim 1$ fm.

The central-depression phenomena of the transition density 
originates from the behavior of the form factor in the range 
of $q^2 \gtrsim 10$ fm$^{-2}$,
and the observed transition form factor~\cite{Kegel} limited to the range 
of $q^2 \leq 5 \,{\rm fm}^{-2}$ cannot provide information about
the central-depression structure at \mbox{$r \lesssim 1 $ fm.}
However, we considered that using the observed form factor 
can be used to derive the transition density $\rho_{\rm tr}(r)$
in the range $r \gtrsim 1$ fm,
as indicated in Fig.~4b by the solid and dashed red lines, 
which is significantly different from the
black and green lines ($r \gtrsim 1$ fm) obtained in the 
literature~\cite{second2004,Bacca2013}.

\vskip 0.3cm
\noindent
{\bf 4. Monopole matrix element \boldmath{$\langle r^2 \rangle_{\rm tr}$} and
transition radius \boldmath{$\mathcal{R}_{\rm tr}$}}
 
As important information on the spatial structure of the second $0^+$ state,
the authors of Ref.~\cite{Kegel} extracted, from the observed form factor,
the monopole matrix element $\langle r^2 \rangle_{\rm tr}$ and 
the transition radius  $\mathcal{R}_{\rm tr}$ defined as
(cf. Eq.~(5) in Ref.~\cite{Kegel})
\begin{eqnarray}
\langle r^n \rangle_{\rm tr} =
 Z \left|\, \!\int r^n  \,\rho_{\rm tr}(r)\, {\rm d}{\bf r}\,\right| \quad 
  (n=2 \,\,{\rm and}\,\,4, Z=2),
  \qquad \mathcal{R}_{\rm tr}^2=
\frac{\langle r^4 \rangle_{\rm tr}}{\langle r^2 \rangle_{\rm tr}},
\end{eqnarray}
which are obtained by   a ${\bf q} \to 0$ expansion of the form factor
\begin{eqnarray}
Z \,|{\widetilde {\cal F}}_{{\rm M}0^+}(q^2)|/q^2
=\tfrac{1}{6} \langle r^2 \rangle_{\rm tr} 
- \tfrac{1}{120}\langle r^4 \rangle_{\rm tr} \,q^2 + {\cal O}(q^4),
\end{eqnarray}
with $e^{i{\bf q}\cdot{\bf r}} \to 1 - \tfrac{1}{6} q^2 r^2 
+ \tfrac{1}{120} q^4 r^4 + \cdot \cdot \cdot $ for the ``monopole'' density 
$\rho_{\rm tr}(r)$ in Eq.~(5).
They determined the numerical values of 
$\langle r^2 \rangle_{\rm tr}$ and $\mathcal{R}_{\rm tr}$,
as indicated by the first  line in Table~I.

We  calculate $\langle r^2 \rangle_{\rm tr}$ and 
$\mathcal{R}_{\rm tr}$ in Eq.~(14) explicitly using the red-solid-line density,
Eq~(8), with no central depression 
and the red-dashed-line density, Eq.(8) plus Eq.~(11), with central depression. 
We obtain, for the  red-dashed-line density,
\begin{eqnarray}
\langle r^2 \rangle_{\rm tr} = 6\, Z\, |\, a_1+a_2+a_3\, | , \qquad
\langle r^4 \rangle_{\rm tr}  = 120\, Z\,|\, a_1 b_1 +a_2 b_2+ a_3 b_3\, |,
\end{eqnarray}
where $a_3$ and $b_3$ are omitted in the case of the red-solid-line density.
The numerical values of $\langle r^2 \rangle_{\rm tr}$ and
$\mathcal{R}_{\rm tr}$   are listed in Table~I. The two  transition densities
reproduce the experimental value; however, this is natural because
the form factors of Eqs.~(2) and (10) are constructed 
to simulate the behavior of the observed form factor (blue line) at $q^2 \sim 0$.
As expected, the central-depression structure is not reflected
in $\langle r^2 \rangle_{\rm tr}$ and 
$\mathcal{R}_{\rm tr}$ (that is, the contributions of $a_3$ and $b_3$ are
negligible in Eq.~(15)). 

\begin{table}[t!]
\caption{ \;\;
Values of $\langle r^2 \rangle_{\rm tr}$ and 
$\mathcal{R}_{\rm tr}$. 
}
\begin{center}
\begin{tabular}{lcccc} 
\hline \hline
\noalign{\vskip 0.1 true cm} 
  & $\quad$ &   $\langle r^2 \rangle_{\rm tr}\, ({\rm fm}^2)$ 
  & $\quad$ &  $\mathcal{R}_{\rm tr}\, ({\rm fm})$ \\
\noalign{\vskip 0.1 true cm} 
\hline
\noalign{\vskip 0.15 true cm} 
Experiment, Ref.~\cite{Kegel} &  &  $1.53 \pm 0.05$ & &   $4.56 \pm 0.15$      \\
\noalign{\vskip 0.15 true cm} 
Red-solid-line density, Eq. (8) &  &  $1.52 $ & &   $4.65$      \\
\noalign{\vskip 0.15 true cm} 
Red-dashed-line density, Eqs.~(8) plus (11) &  &  $1.52 $ & &   $4.65$      \\
\noalign{\vskip 0.15 true cm} 
\hline
\hline
\end{tabular}
\end{center}
\label{table:meen-radius}
\end{table}

\begin{figure} [h!]
\begin{center}
\epsfig{file=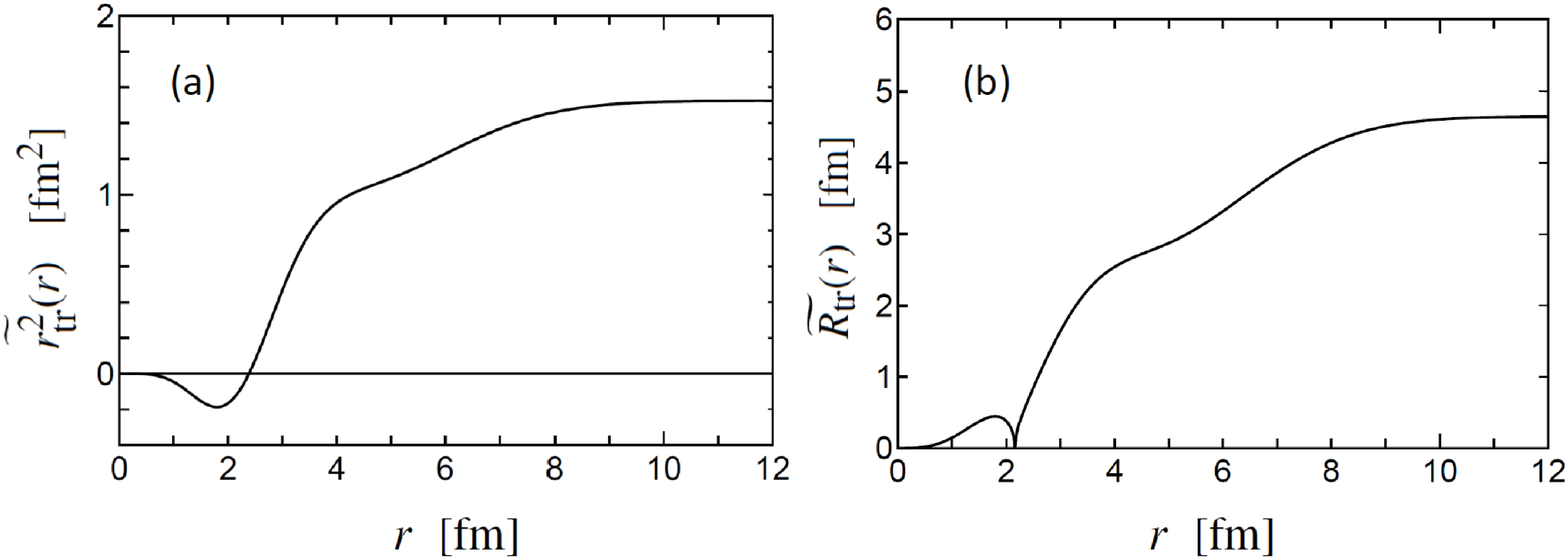,width=15.0cm}
\caption{ 
(a) 
Cumulative monopole matrix element ${\widetilde {r^2_{\rm tr}}}(r)$
and (b) 
Cumulative transition density ${\widetilde {\mathcal{R}_{\rm tr}}}(r)$ 
defined by Eq.~(\ref{eq:cumulative}). 
In each of (a) and (b), the lines for the red-solid-line and 
red-dashed-line densities in Fig.~4b overlap 
within the thickness of the lines.
Here, ${\widetilde {r^2_{\rm tr}}}(\infty) = \langle r^2 \rangle_{\rm tr}
=1.52 \,{\rm fm}^{2}$  and
${\widetilde {\mathcal{R}_{\rm tr}}}(\infty) = \mathcal{R}_{\rm tr}
=4.65 \,{\rm fm}^{-2}$ (cf. Table~I).
} 
\label{fig:cumu-dens-fig5ab}
\end{center}	
\end{figure}

To investigate the range of $r$ that contributes most to  
$\langle r^2 \rangle_{\rm tr}$ and  $\mathcal{R}_{\rm tr}$,
we introduce the {\it cumulative} monopole matrix element ${\widetilde {r^2_{\rm tr}}}(r)$ 
and the {\it cumulative} transition density 
${\widetilde {\mathcal{R}_{\rm tr}}}(r)$ as a function of $r$ by 
\begin{eqnarray}
{\widetilde {r^2_{\rm tr}}}(r) \equiv 4\pi Z \left|  \int_0^{\, r} 
\! {r'}^2 \rho_{\rm tr}({r'})\, r'^2 {\rm d}{r'} \,\right| ,
\qquad  
{\widetilde {\mathcal{R}_{\rm tr}}}(r) \equiv  
\left| \, \frac{4\pi Z }{\langle r^2 \rangle_{\rm tr}}  
\int_0^{\, r} \! r'^4 \rho_{\rm tr}(r')\, r'^2 {\rm d}r'\,\right|^{1/2},
\label{eq:cumulative}
\end{eqnarray}
where ${\widetilde {r^2_{\rm tr}}}(\infty) = \langle r^2 \rangle_{\rm tr}$
and
${\widetilde {\mathcal{R}_{\rm tr}}}(\infty) = \mathcal{R}_{\rm tr}$.
The functions ${\widetilde {r^2_{\rm tr}}}(r)$ and
${\widetilde {\mathcal{R}_{\rm tr}}}(r)$ are shown in Figs.~5a and 5b,
respectively. 
In both cases, the dominant contribution comes from the range of
\mbox{$ r \gtrsim 3$ fm.} Interestingly, this part appears to be minor 
in the region of Fig.~4b for the transition density. 
We understand that we must derive the transition density
$\rho_{\rm tr}(r)$ up to $r\! \sim \!10$ fm so that 
$\langle r^2 \rangle_{\rm tr}$ and $\mathcal{R}_{\rm tr}$ are calculated 
accurately.

\vskip 0.3cm
\noindent
{\bf 5. Discussions}

We have derived  a possible $0^+_1\! \to\! 0_2^+$ transition density 
$\rho_{\rm tr}(r)$ 
for $ r \gtrsim 1 $ fm as in Fig.~4b by the solid red line
except for the inner central-depression region,
utilizing the information obtained from the newly observed high-precision 
transition form factor for \mbox{$0 \leq q^2 \leq 5 \,{\rm fm}^{-2}$~\cite{Kegel}.}
The shape of the transition density is significantly different 
from those obtained theoretically in literature~\cite{second2004,Bacca2013}.

We now discuss a problem calculating the energy of the second $0^+$ state.
Figure~6 \mbox{schematically} illustrates the energy values obtained
using the AV8$'$+central 3N potential~\cite{second2004}
and by the $\chi$EFT and AV18+UIX potentials~\cite{Bacca2013} along with the
threshold energies of the $\,^3{\rm H}+p$, $^3{\rm He}+n$ and their average
$3N+N$ configurations~\cite{Tilley1992}. Note that the observed $0^+_2$ state 
is located between the $^3{\rm H}+p$ and  $^3{\rm He}+n$ thresholds and
only 0.01 MeV above the $3N+N$ threshold 
with a rather narrow  width of $\Gamma_{\rm p} = 0.50$  MeV~\cite{Tilley1992}
 for a $S$-wave resonance.

\begin{figure} [b!]
\begin{center} 
\epsfig{file=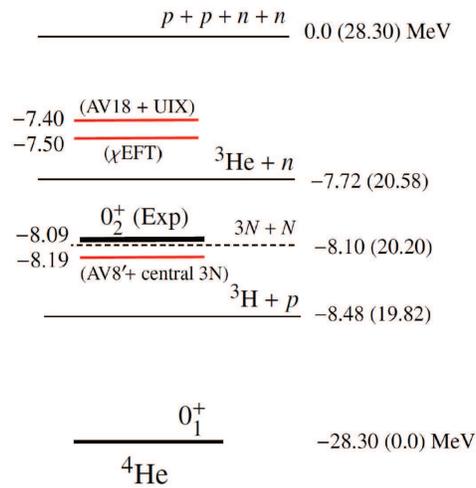,width=6.5cm}
\caption{ Schematic illustration of the $0^+_2$ energy obtained
using the AV8$'$+central 3N potential~Ref.~\cite{second2004}
and by the  AV18+UIX and $\chi$EFT potentials~\cite{Bacca2013} along with the
threshold energies of the $^3{\rm H}+p$, $^3{\rm He}+n$, their averaged
$3N+N$, and $p+p+n+n$  configurations~\cite{Tilley1992}.
} 
\label{fig:FF-fig2}
\end{center}	
\end{figure}

In Ref.~\cite{second2004} which took the isospin-formalism,
the $0^+_2, T\!=\!0$ state was obtained 
only 0.1 MeV below the observed level as 
a {\it bound} state measured from the $3N+N$ threshold.
\mbox{On the other hand,} in Ref.~\cite{Bacca2013},
the energy of the $0^+_2$ state obtained 
as a resonance with the two realistic interactions 
is $\sim \! 0.7$ MeV above the observed $0^+_2$ level and 
even above the $^3{\rm He}+n$ threshold.
As mentioned  by Bacca {\it et al.}~\cite{Bacca2013} 
and Epelbaum~\cite{Epelbaum}, 
it is possible that the improper theoretical resonance position
affects the form factor result.
As pointed  by Horiuchi and Suzuki (cf. last paragraph of Sec.~IIIB
of Ref.~\cite{Horiuchi}),
the observed $0^+_2$ state is considered  a Feshbach resonance
embedded in the $^3{\rm H}+p$ continuum as a {\it bound} state with respect to
the  $^3{\rm He}+n$ threshold.

In the opinion of the present author, 
\mbox{a possible} strategy to attack the $\alpha$-particle monopole puzzle 
using fully realistic interactions is as follows:
(i) Solve  the $0^+_2$ state using the bound-state \mbox{approximation} to search 
interaction parameters that reproduce the $0^+_2$ energy as well as
the binding energies of $^3{\rm H}$, $^3{\rm He}$ and $^4{\rm He}$,
\mbox{(ii) calculate} the transition density and form factor, 
and then \mbox{(iii) solve} the $^3{\rm H}+p$ scattering
to derive the wave function and width of the Feshbach resonance
as well as the final results for the transition density and form factor. 

 As mentioned in Sec.~1, authors of Ref.~\cite{Kegel} claimed 
the following: the large difference between the calculated form factors does not 
stem from numerical methods but from the Hamiltonian.
However, the comparison between the methods 
was performed only for the calculation based
on the {\it bound-state} approximation.  
The calculation of the $0^+_2$ state as a {\it resonance} 
was performed using only the LIT approach.
For example, it would be desirable to examine this problem
by the comparison with the form factor results produced by using
any explicit four-body
$^3{\rm H}+p$ scattering calculation such as Ref.~\cite{Viviani2020}. 
 

\vskip 0.3cm
\noindent
{\bf  Acknowledgements}

The author would like to thank Dr. S. Kegel for providing with the numerical
values of the experimental data of Ref.~\cite{Kegel} and
valuable discussions on the data.
This work is supported by 
the Grant-in-Aid for Scientific Research on Innovative Areas, 
``Toward new frontiers: Encounter and synergy of state-of-the-art 
astronomical detectors and exotic quantum beams", Grant Number 18H05461.



\begin{thebibliography}{99}

\bibitem{Kama01} H.\ Kamada, A.\ Nogga, W.\ Gl\"{o}ckle, E.\ Hiyama,
   M.\ Kamimura, K.\ Varga, Y.\ Suzuki, M.\ Viviani, A.\ Kievsky,
   S.\ Rosati, J.\ Carlson, S.\ C.\ Pieper, R.\ B.\ Wiringa, P.\
   Navratil, B.\ R.\ Barrett, N.\ Barnea, W.\ Leidemann, and G.\
   Orlandini, Phys. Rev. C {\bf 64}, 044001 (2001).

\bibitem{AV8P97} B.\ S.\ Pudliner, V.\ R.\ Pandharipande, J.\
   Carlson, S.\ C.\ Pieper, and R.\ B.\ Wiringa, Phys.\ Rev.\
   C {\bf 56}, 1720 (1997).

\bibitem{Kamimura88} M.\ Kamimura, Phys.\ Rev.\ A{\bf 38}, 621 (1988).

\bibitem{Kameyama89} H.\ Kameyama, M.\ Kamimura and Y.\ Fukushima,
   Phys.\ Rev.\ C {\bf 40}, 974 (1989).

\bibitem{Hiyama03} E.\ Hiyama, Y.\ Kino and M.\ Kamimura,
   Prog.\ Part.\ Nucl.\ Phys.\ {\bf 51}, 223 (2003).

\bibitem{second2004} E. Hiyama, B.F. Gibson, and M. Kamimura,
Phys. Rev. C {\bf 70}, 031001(R) (2004).

\bibitem{Fro68} R.\ F.\ Frosch, R.\ E.\ Rand, H.\ Crannell, J.\ S.\
   McCarthy, L.\ R.\ Suelzle, and M.\ R.\ Yearian, Nucl.\ Phys.\
   A{\bf 110}, 657 (1968).

\bibitem{Wal70} Th.\ Walcher, Phys.\ Lett.\ B, {\bf 31}, 442 (1970).

\bibitem{Koe83} G.\ Koebschall, C.\ Ottermann, K.\ Maurer,
   K.\ Roehrich, Ch.\ Schmitt and V.\ H.\ Walther, Nucl.\ Phys.\ 
   A{\bf 405}, 648 (1983).

\bibitem{Horiuchi} W. Horiuchi and Y. Suzuki, Phys. Rev. C {\bf 78}, 034305 (2008).

\bibitem{stochas-1} Y. Suzuki and K. Varga, {\it Stochastic Variational Approach to
Quantum Mechanical Few-Body Problems} (Springer-Verlag,
Berlin, 1998).

\bibitem{stochas-2} K. Varga and Y. Suzuki, Phys. Rev. C 52, 2885 (1995).


\bibitem{Bacca2013} S. Bacca,  N. Barnea,  W. Leidemann, and G. Orlandini, 
Phys. Rev. Lett. {\bf 110}, 042503, (2013). 

\bibitem{Bacca2015} S. Bacca,  N. Barnea,  W. Leidemann, and G. Orlandini, 
Phys. Rev. C {\bf 91}, 024303, (2015). 

\bibitem{EIHH1997} N. Barnea and A. Novoselsky, Phys. Rev. A {\bf 57}, 48 (1998).

\bibitem{EIHH2001} N. Barnea, W. Leidemann, and G. Orlandini, Phys. Rev. C
{\bf 61}, 054001 (2000). 

\bibitem{LIT1994} V.D. Efros, W. Leidemann, and G. Orlandini, Phys. Lett.
B {\bf 338}, 130 (1994).

\bibitem{LIT2007} V.D. Efros, W. Leidemann, G. Orlandini, and N. Barnea,
J. Phys. G {\bf 34}, R459 (2007).

\bibitem{AV18} R.B. Wiringa, V.G.J. Stoks, and R. Schiavilla, Phys. Rev.
C {\bf 51}, 38 (1995).

\bibitem{UIX} B.S. Pudliner, V. Pandharipande, J. Carlson, and 
R. Wiringa, Phys. Rev. Lett. {\bf 74}, 4396 (1995).

\bibitem{EFT-NN} D. R. Entem and R. Machleidt, Phys. Rev. C {\bf 68}, 
041001(R) (2003).

\bibitem{EFT-3FN-1} P. Navratil, Few-Body Syst. {\bf 41}, 117 (2007).

\bibitem{EFT-3FN-2} D. Gazit, S. Quaglioni, and P. Navratil, 
Phys. Rev. Lett. {\bf 103}, 102502 (2009).

\bibitem{Kegel} S. Kegel {\it et al.},
Phys. Rev. Lett. {\bf 130}, 152502, (2023). 

\bibitem{Epelbaum} E. Epelbaum, Physics {\bf 16}, 58 (2023).

\bibitem{Tilley1992} D.R. Tilley, H.R. Weller, and G.M.Hale,
Nucl. Phys. A {\bf 541}, 1 (1992).

\bibitem{Viviani2020} M. Viviani, L. Girlanda, A. Kievsky, and L.E. Marcucci,
Phys. Rev. C {\bf 102}, 034007 (2020).

\end{thebibliography}
\end{document}